\documentclass[showpacs,showkeys,preprintnumbers,amsmath,amssymb,nofootinbib]{revtex4}

\usepackage{graphicx}
\usepackage{dcolumn}
\usepackage{bm}
\usepackage{multirow}
\usepackage{subfigure}

\begin{document}

\title{A comprehensive revisit of the $\rho$ meson with improved Monte-Carlo based QCD sum rules}

\author{Qi-Nan Wang$^1$}
\author{Zhu-Feng Zhang$^{1,2}$}
\author{T. G. Steele$^2$}
\author{Hong-Ying Jin$^3$}
\author{Zhuo-Ran Huang$^{2,3}$}

\affiliation{
$^1$Physics Department, Ningbo University, Zhejiang Province, 315211, P. R. China
\\
$^2$Department of Physics and Engineering Physics, University of Saskatchewan, Saskatoon,
Saskatchewan, S7N 5E2, Canada
\\
$^3$Zhejiang Institute of Modern Physics, Zhejiang University, Zhejiang Province, 310027, P. R. China
}

\begin{abstract}
We improve the Monte-Carlo based QCD sum rules by introducing the rigorous H\"older-inequality-determined
sum rule window and a Breit-Wigner type parametrization for the phenomenological spectral function. In this
improved sum rule analysis methodology, the sum rule analysis window can be determined without any assumptions
on OPE convergence or the QCD continuum.
Therefore an unbiased prediction can be obtained for the phenomenological parameters (the hadronic mass and width etc.). 
We test the new approach in the $\rho$ meson channel with re-examination
and inclusion of $\alpha_s$ corrections to dimension-4 condensates in the OPE. We obtain results highly consistent
 with  experimental values. We also discuss the possible extension of this method to some other channels.
\end{abstract}



\maketitle

\section{Introduction}

QCD sum rule (QCDSR) is an important nonperturbative method in hadronic physics, introduced by
Shifman et al. (SVZ) \cite{SVZ1,SVZ2}. Due to its QCD based nature and minimally model dependent property,
this semi-analytic approach has become a powerful weapon in the extensive studies on phenomenological
hadronic properties.\footnote{For reviews, see \cite{RRY,Narison:2014wqa,Narison:2010wb}. } Although
fruitful results have been achieved within the framework of QCD sum rules, doubts on the predictive
capability of this approach have never disappeared since in the original Laplace/Borel sum rules,
the two external free parameters, i.e., Borel  parameters $\tau$ and continuum thresholds $s_0$,
cannot be rigorously constrained. Intuitive  choices of ranges for ($\tau$, $s_0$) cannot completely exclude
subjective factors in determination of the phenomenological outputs.

Two main directions have been developed to  address the above issues. First, the
sum rule stability criteria \cite{Narison:2014wqa,Narison:2010wb} in Laplace sum rules (LSR),
systematically developed by Narison, have been widely and successfully used to study QCD phenomenological
properties.\footnote{For recent applications of the sum rule stability criteria, see
\cite{Albuquerque:2016znh,Albuquerque:2016nlw,Huang:2016rro,Huang:2016upt,Narison:2015nxh,Narison:2014vka}.}
In the widely used  ``single narrow resonance minimal duality ansatz'' where the hadronic widths are not involved,
the mass curves present stability (if any) in $\tau$ or $s_0$ at the extremums, where the mass depends weakly on
($\tau$, $s_0$) and thus optimal predictions can be achieved if validity of operator product expansion (OPE) truncation and pole contribution
dominance are simultaneously ensured. In the cases where the $s_0$-stability is absent, the finite energy sum rules
(FESR) \cite{Shankar:1977ap} can complement the systematics of the method \cite{Narison:2009vj,Matheus:2004qq,Huang:2016rro}.
These approaches require an assessment of whether the optimized values for $\tau$ and $s_0$ are in the sum rule region of validity.

The Monte-Carlo based weighted-least-squares fitting procedure is another quite successful approach to allow reliable
sum rule predictions \cite{Leinweber:1995fn,Lee:1996dc,Lee:1997ix,Lee:1997ne,Wang:2008vg,Erkol:2008gp,Zhang:2013rya,Huang:2014hya}.
As originally suggested by SVZ \cite{SVZ1,SVZ2}, the sum rules should be analyzed in a certain range of the Borel parameter,
called the ``sum rule window'', in which the OPE series reach convergence and the continuum contribution are suppressed. Directly
following this discussion, Leinweber introduced the Monte-Carlo based weighted-least-squares fitting procedure \cite{Leinweber:1995fn}
in the sum rule numerical analysis. Within this approach, phenomenological outputs, such as the mass, the decay constant, and the
continuum threshold etc., can be obtained by minimizing the weighted-$\chi^2$ function between the OPE and phenomenological
expressions of the correlation function within the sum rule window. This method has some obvious advantages:
\begin{itemize}
\item the continuum
threshold $s_0$ can be determined from the fitting procedure;
\item different parametrization models for the hadronic spectral function can be dealt with therefore
it is possible to obtain predictions both for hadronic mass and width with an appropriate spectral density;
\item uncertainty analyses and dependence of the outputs on the phenomenological input parameters can be provided via the Monte-Carlo
based numerical procedure.
\end{itemize}

However, as in the original (as well as many common uses of) QCD sum rules, the sum rule window cannot
be rigorously constrained.\footnote{A remedial strategy is to determine to optimal outputs by observing the variation of the outputs
on the change of the conditions imposed on the sum rule window. If the outputs are not sensitive to the variation of the range of the
window, reliable predictions can be obtained by demanding the OPE converge in a proper trend, as did in \cite{Huang:2014hya} for the
exotic hybrids.} The usual imposed conditions, the pole contribution $>$ 50$\%$ of total phenomenological expression and the highest
dimension operator (HDO) contribution $<$ 10$\%$ of the total OPE are based upon assumptions, and cannot always be ensured especially
for multi-quark state sum rules where the OPE convergence is slower, which makes artificial adjustments of these constraints inevitable.

The H\"older inequalities for QCD sum rules, obtained from the requirement that the imaginary part of a
correlation function should be positive because of its relation to a hadronic spectral function
(e.g., physical cross sections for electromagnetic currents), were first introduced as
fundamental constraints on the sum rules by Benmerrouche et al. \cite{Benmerrouche:1995qa}, then were used
in many research works on QCD sum rules, including constraints on the QCD sum rule window
\cite{Steele:1996np,Steele:1998ry,Shi:1999hm,Kleiv:2013dta,Chen:2016jxd}. However,
these inequalities have mainly been applied in analyses of the simplified
``single narrow resonance minimal duality ansatz''.

In fact, most hadrons listed in Review of Particle Physics \cite{Olive:2016xmw} are resonances rather than
stable particles, thus the mass and width are equally important in describing the properties of these hadrons,
and equal attention should be paid to them. Some researchers replace the delta function in spectral density directly with
a Breit-Wigner (BW) type function in order to obtain information for the hadron width in QCD sum rules
\cite{Leupold:1997dg,Lee:2008tz,Erkol:2008gp}. In this paper, we will introduce a BW type parametrization for the
phenomenological spectral function that satisfies the low-energy theorem for the form factor, and apply
the rigorous H\"older-inequality-determined sum rule window in the Monte-Carlo
based fitting procedure. We will reanalyze the phenomenological properties of the $\rho$ meson with this new systematic
numerical procedure. A comprehensive study will be presented, including the explicit reexamination of some $\alpha_s$
corrections in the OPE\footnote{These radiative corrections to dimension-4 quark and gluon condensates are complicated
 calculations but may play an important role in determining the range of $\tau$ where the H\"older inequalities are
satisfied. The previous results of these radiative corrections obtained using a projector method in \cite{Surguladze:1990sp} were not used in the existing
numerical analyses \cite{Leinweber:1995fn,Leupold:1997dg}, motivating our independent calculation.}
and uncertainty analysis for the fitted results.
Extensions of our analysis to include a smooth transition to the QCD continuum are also presented.
We will conclude the paper by summarizing our calculation and analysis
and discussing the possible extension of our approach.

\section{Operator product expansion for correlation function for vector $I=1$ current}

The starting point of QCD sum rules is to calculate the correlation function for a specific current. In this paper, we consider the
light vector current with isospin 1, i.e., $j_\mu=\bar u \gamma_\mu d$ (and $\frac{1}{\sqrt{2}}(\bar u\gamma_\mu u-\bar d\gamma_\mu d)$,
$\bar d\gamma_\mu u$), which has the same quantum numbers as the $\rho$ meson. Because of
conservation of the vector current, the correlation function for current $j_\mu$ can be written as
\begin{equation}
\label{eq:correlator}
\Pi_{\mu\nu}(q^{2})=i\int d^4x \,e^{i qx}\langle 0|T{j_{\mu}(x)j^\dagger_{\nu}(0)}|0\rangle =(q_{\mu}q_{\nu}-q^2g_{\mu\nu})\Pi(q^{2}),
\end{equation}
where $\Pi(q^2)$ is a Lorentz scalar function which can be calculated theoretically using operator product expansion methods
\cite{Novikov:1983gd}. The well-confirmed and widely used QCD expression of  $\Pi (q^2)$, including leading-order (LO)
and next-to-leading order (NLO) perturbative terms and LO non-perturbative terms up to dimension-6 condensates reads \cite{SVZ1,SVZ2}:
\begin{equation}
\label{eq:ope0}
\Pi^{\textrm{(OPE)}}_0(q^{2})=-\frac{1}{4\pi^{2}}\left(1+\frac{\alpha_{s}}{\pi}\right)\ln \left(\frac{-q^{2}}{\mu^{2}}\right)+
\frac{1}{12\pi}\frac{\langle\alpha_{s}G^2\rangle}{(q^{2})^{2}}+\frac{2\langle m_{q} \bar{q}q\rangle}{(q^{2})^{2}}+\frac{224\pi}{81}
\frac{\kappa\alpha_{s} \langle \bar{q}q\rangle^{2}}{(q^{2})^{3}},
\end{equation}
where $\kappa$ is the factorization violation factor which parameterizes the deviation of the four-quark condensate
from a product of two-quark condensates.

In addition to these terms, some researchers have
also calculated the NLO corrections in $\alpha_s$ to dimension-4 condensates \cite{Chetyrkin:1985kn,Surguladze:1990sp}, which however
are not included in the existing sum rule numerical analyses for $\rho$ meson \cite{Leinweber:1995fn,Leupold:1997dg}. Although the
results in \cite{Chetyrkin:1985kn,Surguladze:1990sp} suggest these corrections are relatively small, they can play an
important role in determining the sum rule window from the H\"older inequalities, motivating our
independent calculation with a different method. Thus, we calculate these contributions using the external field method
in Feynman gauge\footnote{Similar calculations can be seen in \cite{Jin:2002rw,Jin:2000ek} for the ($1^{-+}$,$0^{++}$) light
hybrid states.} \cite{Abbott:1980hw,Novikov:1983gd} to
examine the previous results obtained in \cite{Chetyrkin:1985kn,Surguladze:1990sp}. Our results read
\begin{equation}
\label{eq:alphacor}
\Pi^{\alpha}_{d=4}(q^2)=\frac{1}{(q^{2})^{2}}\frac{7}{72\pi}\frac{\alpha_{s}}{\pi}\langle\alpha_{s}G^2\rangle
+\frac{1}{(q^{2})^{2}}\frac{2}{3}\frac{\alpha_{s}}{\pi}\langle m_{q}\bar{q}q\rangle,
\end{equation}
which is consistent with the previous results in Ref.~\cite{Chetyrkin:1985kn,Surguladze:1990sp}. The whole calculation is
much more complicated than the LO calculation, thus for clarity, we give all the related Feynman diagrams and explicit
results of our calculation in the Appendix.

In the following section, we will use
\begin{equation}
\Pi^{\textrm{(OPE)}}(q^2)=\Pi^{\textrm{(OPE)}}_0(q^{2})+\Pi^{\alpha}_{d=4}(q^2)
\end{equation}
 to reanalyze the sum rules for the $\rho$ meson, and discuss the effects of these $\alpha_s$ corrections.

\section{Monte-Carlo based QCD sum rules for $\rho$ meson}

To obtain a QCD sum rule which can be used to predict hadronic properties, we first need to Borel-transform the theoretical representation
of the correlation function, i.e., $\Pi^{\textrm{(OPE)}}(q^2)$. After some calculations, we obtain
\begin{equation}
\label{eq:theoretical}
\begin{split}
R^{\textrm{(OPE)}}(\tau)=&\frac{1}{\tau}\hat B \Pi^{\textrm{(OPE)}}(q^2)=\frac{1}{4\pi^2}
\left(1+\frac{\alpha_s(1/\tau)}{\pi}\right)\frac{1}{\tau}+\frac{\langle\alpha_s G^2\rangle}{12\pi}
\left(1+\frac{7}{6}\frac{\alpha_s(1/\tau)}{\pi}\right) \tau \\
&+ 2\langle m_q\bar qq\rangle \left(1+\frac{1}{3}\frac{\alpha_s(1/\tau)}{\pi}\right) \tau-\frac{112}{81}
\pi \kappa \alpha_s\langle \bar qq\rangle^2 \left[\frac{\alpha_s(1/\tau)}{\alpha_s(\mu^2_0)}\right]^{1/9}  \tau^2,
\end{split}
\end{equation}
where $\hat B$ is the Borel transformation operator and $\alpha_s(1/\tau)=4\pi/(9\ln(1/(\tau \Lambda^2_{\textrm{QCD}})))$
is the running coupling constant for three flavors at scale $1/\sqrt \tau$.
We have considered the renormalization-group (RG) improvement of the sum rules \cite{Narison:1981ts} and anomalous dimensions
for condensates \cite{SVZ1,SVZ2} in Eq.~\eqref{eq:theoretical}, where $\mu_0$ is the renormalization scale for condensates.

Eq.~\eqref{eq:theoretical} provides the theoretical representation of Borel-transformed correlation function, but in order to obtain
a QCD sum rule, we still need the phenomenological representation obtained by constructing a phenomenological spectral density model.

If we insert a complete set of one-particle states $\int\frac{d^3 {\bm k}}{(2\pi)^3 2 E_{\bm{k}}} |\rho({k})\rangle\langle\rho({k})|+\textrm{``other states''}$
into correlation function, we will reach a phenomenological spectral density with a delta function $\delta(s-m_\rho^2)$,
which is widely used in traditional QCD sum rules. However, the ground state of light vector $I=1$ meson, i.e., $\rho$ meson, is a resonance which
is far away from a stable particle, thus it's more appropriate to insert two-particles intermediate states into the correlation function since the two-pion
decay mode is the dominant decay mode for $\rho$ meson.
By inserting two-pion intermediate states into Eq.~\eqref{eq:correlator}, e.g., inserting $\int\frac{d^3 {\bm k_1}}{(2\pi)^3 2 E_{\bm{k_1}}}
\frac{d^3{\bm k_2}}{(2\pi)^3 2 E_{\bm{k_2}}} |\pi^+(k_1)\pi^0(k_2)\rangle \langle \pi^+(k_1)\pi^0(k_2)| + \textrm{``other intermediate states''}$
for correlation function of
current $j_\mu=\bar d \gamma_\mu u$, and using the Cutkosky's cutting rules \cite{Cutkosky:1960sp},
the phenomenological expression for ${\rm Im}\Pi(s)$ can be found \cite{DMP}:
\begin{equation}
{\rm Im}\Pi^{\textrm{(phen)}}(s)=\frac{1}{24\pi}\left[1-\frac{4m_\pi^2}{s}\right]^{3/2}|F_\pi(s)|^2+\textrm{contributions from excited states and continuum},
\end{equation}
where $m_\pi$ is the mass of pion, and
\begin{equation}
\langle 0|j_\mu(0)|\pi^+(k_1)\pi^0(k_2)\rangle =\sqrt{2} (k_1-k_2)_\mu F_\pi((k_1+k_2)^2)
\end{equation}
has been used, where $F_\pi(s)$ is the electromagnetic form factor which is normalized as $F_\pi(0)=1$.
Furthermore, since the main contribution to $F_\pi(s)$ comes from the $\rho$ meson, $F_\pi(s)$ should have a pole at $s=m_\rho^2-i m_\rho \Gamma_\rho$,
where $m_\rho$ and $\Gamma_\rho$ are the mass and width of $\rho$ meson respectively.

Following Ref.~\cite{Dominguez:1986td} we use a Breit-Wigner form function to construct a model for the form factor as follows
\begin{equation}
|F_\pi(s)|^2=\frac{m_\rho^4+m_\rho^2\Gamma_\rho^2}{(s-m_\rho^2)^2+m_\rho^2\Gamma_\rho^2},
\end{equation}
which meets the above requirements (including the low-energy theorem $F_\pi(0)=1$),
and is more simple than the Gounaris-Sakurai parameterized form factor used in Ref.~\cite{DMP} and
the parameterized form factor used in Ref.~\cite{Martinovic:1990hy,Martinovic:1991ss}.
As outlined below, the  phenomenological spectral function normalization constrained by a low-energy theorem enters our analysis in a meaningful way because we
work directly with the Borel-transformed correlation function $R(\tau)$ rather than the 
typical sum rule approach which uses normalization-independent ratios of sum rules.
For the excited states and continuum (ESC) contributions in
the spectral density, we still use the same model as traditional
QCD sum rules, i.e., we use a spectral density as follows in this paper:
\begin{equation}
\begin{split}
\label{eq:model}
\frac{1}{\pi}{\rm Im}\Pi^{{\rm (phen)}}(s)=&\frac{1}{24\pi^2}|F_\pi(s)|^2+\frac{1}{\pi}{\rm Im}\Pi^{{\rm (OPE)}}(s)\theta(s-s_0)\\
=&\frac{1}{24\pi^2}\frac{m_\rho^4+m_\rho^2\Gamma_\rho^2}{(s-m_\rho^2)^2+m_\rho^2\Gamma_\rho^2}
+\frac{1}{4\pi^2}\left(1+\frac{\alpha_s}{\pi}\right)\theta(s-s_0),
\end{split}
\end{equation}
where $s_0$ is the continuum threshold separating the contributions from excited states, and we have omitted the small mass of the
pion.

By using the dispersion relation, we can obtain the phenomenological representation for Borel-transformed correlation function, which has a following form:
\begin{equation}
\label{eq:phen}
R^{\textrm{(phen)}}(\tau,s_0,m_\rho,\Gamma_\rho)=\frac{1}{\pi}\int_0^\infty e^{-s\tau}{\rm Im}\Pi^{\textrm{(phen)}}(s) \, ds.
\end{equation}
Then the master equation for QCD sum rule can be obtained by demanding the equivalence between Eq.~\eqref{eq:theoretical} and \eqref{eq:phen}:
\begin{equation}
\label{eq:master}
R^{\textrm{(OPE)}}(\tau)=R^{\textrm{(phen)}}(\tau,s_0,m_\rho,\Gamma_\rho),
\end{equation}
which can be used to obtain the predictions for $s_0$, $m_\rho$ and $\Gamma_\rho$.

Obviously, because of the truncation of OPE and the simplicity of the phenomenological spectral density, Eq.~\eqref{eq:master} can not
be valid for all $\tau$, thus one requires a sum rule window in which the validity of the master equation can be established. Usually, a
sum rule window is determined by demanding the highest dimension operator contributions be limited to 10\% of total OPE contributions
and the continuum contributions be limited to 50\% of total phenomenological contributions
\cite{Leinweber:1995fn,Leupold:1997dg,Erkol:2008gp}. However, the setting of percentage 10\% and 50\% is
somewhat arbitrary.

Benmerrouche et al. presented a method based on the H\"older inequality which provides fundamental constraints on QCD sum
rules \cite{Benmerrouche:1995qa}.  The H\"older inequality for integrals defined over a measure $d\mu$ \cite{Beckenbach:1961,Berberian:1965} is
\begin{equation}
\label{eq:hd0}
\left|\int_{s_1}^{s_2} f(s) g(s)d \mu\right|\leq \left(\int_{s_1}^{s_2}|f(s)|^p d\mu\right)^{1/p} \left(\int_{s_1}^{s_2}|g(s)|^q d\mu\right)^{1/q},
\end{equation}
where $1/p+1/q=1$ and $p$, $q\geq 1$. Since $\textrm{Im} \Pi(s)$ is positive because of its relation to
spectral functions (and in some cases to physical cross
sections), it can serve as the measure $d\mu=\textrm{Im}\Pi(s) ds$ in Eq.~\eqref{eq:hd0}. By placing the excited states and continuum 
contributions on the OPE side, we obtain
\begin{equation}
\begin{split}
R^{\textrm{(OPE-ESC)}}(\tau, s_0)&\equiv
R^{\textrm{(OPE)}}(\tau)-R^{\textrm{(ESC)}}(\tau,s_0)\\
&=R^{\textrm{(OPE)}}(\tau)-\frac{1}{\pi}\int_{s_0}^\infty e^{-s\tau} \frac{1}{4\pi}\left(1+\frac{\alpha_s}{\pi}\right)\,ds
=\frac{1}{\pi}\int_{0}^{s_0} e^{-s\tau} \textrm{Im}\Pi^{\textrm{(OPE)}}(s)\,ds,
\end{split}
\end{equation}
then the H\"older inequality for QCD sum rules can be written as \cite{Benmerrouche:1995qa}
\begin{equation}
\label{eq:holder}
R^{\textrm{(OPE-ESC)}}( \omega \tau_1+(1-\omega)\tau_2, s_0)\leq \left[R^{\textrm{(OPE-ESC)}}(\tau_1,s_0)\right]^\omega
\left[R^{\textrm{(OPE-ESC)}}(\tau_2,s_0)\right]^{1-\omega},
\end{equation}
where $f(s)=e^{-\omega s\tau_1}$ and $g(s)=e^{-(1-\omega)s \tau_2}$ were used. The parameter $\omega$ is defined as $\omega=1/p$,
which satisfies $0\leq \omega\leq 1$. For parameters $\tau_1$ and $\tau_2$ we demand $\tau_1<\tau_2$. Following Ref.~\cite{Benmerrouche:1995qa},
we will perform a local analysis on Eq.~\eqref{eq:holder} with $\tau_2-\tau_1=\delta \tau=0.01\,\textrm{GeV}^{-2}$.
The only starting point of this inequality is that $\textrm{Im}\Pi(s)$ should be positive because of its relation to physical
spectral functions (including cross sections in the case of electromagnetic currents),
thus Eq.~\eqref{eq:holder} must be satisfied if sum rules are to consistently describe integrated physical spectral functions. The
H\"older inequality can be used to check if a sum rule window is reliable \cite{Benmerrouche:1995qa,Steele:1996np}, or to give extra constraints on some
physical quantities \cite{Steele:1998ry}. It can also be used to determinate the QCD sum rule window \cite{Shi:1999hm,Kleiv:2013dta,Chen:2016jxd}.
In this paper, we will use Eq.~\eqref{eq:holder} as the only constraint on the QCD sum rule window: by choosing the maximally allowed region
of the H\"older inequality, we will determinate a sum rule window directly.

In order to match the two sides of master equation \eqref{eq:master} in the sum rule window, a weighted-least-squares
method \cite{Leinweber:1995fn} will be used in this paper. By randomly generating 200 sets of Gaussian distributed
phenomenological input parameters with given uncertainties at
$\tau_j=\tau_\textrm{min}+(\tau_\textrm{max}-\tau_\textrm{min})\times(j-1)/(n_B-1)$, where $n_B=21$, we can estimate the
standard deviation $\sigma_{\textrm{OPE}}(\tau_j)$ for $R^{\textrm{(OPE)}}(\tau_j)$. Then, the phenomenological output parameters
$s_0$, $m_\rho$ and $\Gamma_\rho$ can be obtained by minimizing
\begin{equation}
\label{eq:chi2}
\chi^2=\sum_{j=1}^{n_B}\frac{(R^{\textrm{(OPE)}}(\tau_j)-R^{\textrm{(phen)}}(\tau_j,s_0,m_\rho,\Gamma_\rho))^2}{\sigma_{\textrm{OPE}}^2(\tau_j)}.
\end{equation}
After obtaining $\sigma_{\textrm{OPE}}$, 2000 sets of Gaussian distributed input parameters with same given uncertainties will be
generated, we will minimize $\chi^2$ to obtain a set of phenomenological output parameters for each set and the uncertainties
of $s_0$, $m_\rho$ and $\Gamma_\rho$ can be estimated based on these results.

\section{Numerical results}

In our numerical analysis, we use the central values of input parameters (at $\mu_0=1\,\textrm{GeV}$) from a recent review
article of QCD sum rules \cite{Narison:2014wqa}. These values read
\begin{equation}
\label{eq:input}
\Lambda_{\textrm{QCD}}=0.353\,\textrm{GeV},~\langle \alpha_s G^2\rangle =0.07\,\textrm{GeV}^4,~
\langle m_q\bar qq\rangle =0.007\times(-0.246)^3\,\textrm{GeV}^4,~\kappa \alpha_s\langle \bar qq\rangle^2=\kappa\times 1.49\times 10^{-4}\textrm{GeV}^6,
\end{equation}
where the factor $\kappa$ indicates the violation of factorization hypothesis in estimating the dimension-6 quark condensates.
The size of $\kappa$ have been observed in different channels to be 2--4 \cite{Chung:1984gr,Narison:1995jr,Narison:2009vy}, which
is the scale we will consider in our analysis.

\begin{figure}[htbp]
\centering
\subfigure[~$\kappa=2$]{\label{fig:sr1}\includegraphics[scale=0.9]{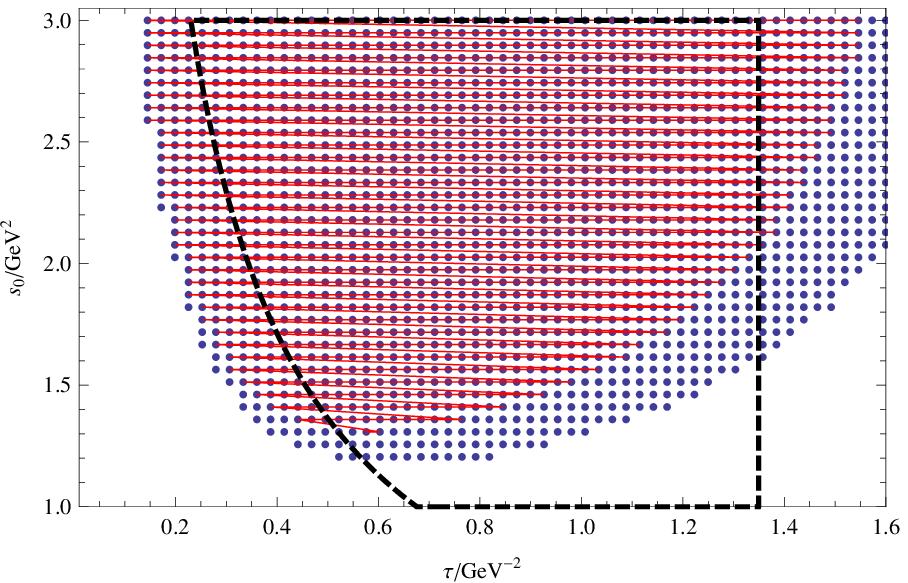}}
\subfigure[~$\kappa=3$]{\label{fig:sr2}\includegraphics[scale=0.9]{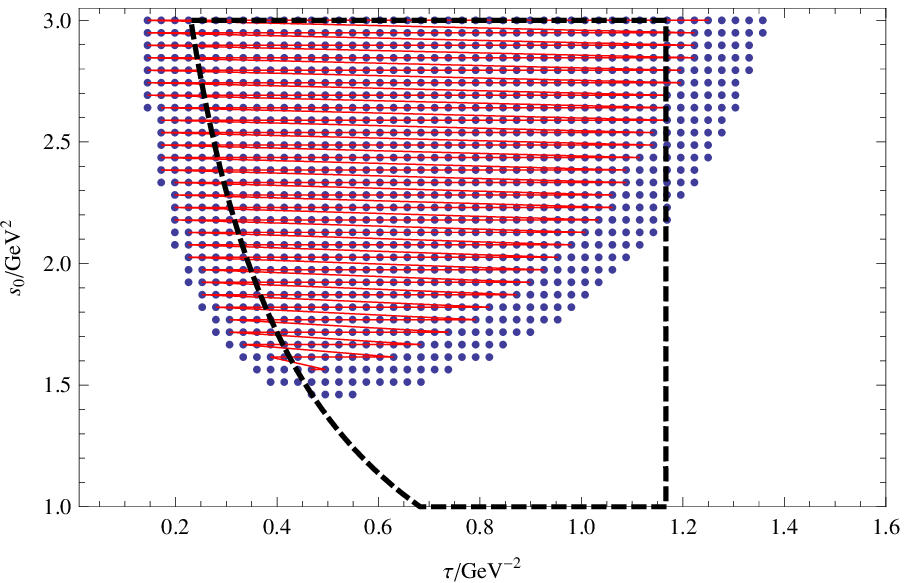}}
\caption{\label{fig:srwindow} The region allowed by the H\"older inequality \eqref{eq:holder} for $\kappa=2$ (a) and $\kappa=3$ (b).
The region with (blue) dot or (red) line is allowed for
sum rule with or without $\alpha_s$ corrections to $\langle \alpha_sG^2\rangle$ and $\langle m_q\bar qq\rangle$
respectively. The region surrounded by the (black) dashed line is obtained by the requirement of HDO/OPE $<$ 10\% and ESC/total $<$ 50\%
with $\alpha_s$ corrections to $\langle \alpha_sG^2\rangle$ and $\langle m_q\bar qq\rangle$. }
\end{figure}

Before fitting the two sides of the master equation \eqref{eq:master}, we should determinate the sum rule window first. In
FIG.~\ref{fig:sr1} and \ref{fig:sr2}, we plot the region which is allowed by the H\"older inequality \eqref{eq:holder} with
$\kappa=2,3$ respectively. For larger factorization violation factor $\kappa$, the allowed region shrinks, however, the main
characteristics of the allowed region, such as the existence of a lower bound on $s_0$, and the bound on $\tau$, will persist for
any value of $\kappa$ from 2 to 4. From FIG.~\ref{fig:sr1} and \ref{fig:sr2}, we find that the $\alpha_s$ corrections to
$\langle\alpha_s G^2\rangle$ and $\langle m_q\bar qq\rangle$ extend the allowed region to a higher $\tau$ region and lower
$s_0$ region. This extension is more obvious in the case with a small $\kappa$. Thus according to the H\"older inequality,
the $\alpha_s$ corrections to dimension-4 operator condensates extend the validity of the sum rule master equation. As a
comparison, we also plot the region obtained by demanding HDO/OPE $<$ 10\% and ESC/total $<$ 50\% with $s_0>1\,\textrm{GeV}^2$.
However, because of the small order of magnitude, the $\alpha_s$ corrections to dimension-4 operators seem to not have important
effects in the determination of the sum rule window from 50\%-10\% method. Whether these corrections are included or not, the allowed
region does not change a lot, thus we do not plot the region for which such corrections are not considered. From FIG.~\ref{fig:srwindow},
we also observe that the constraint on the upper bound of $\tau$ from the H\"older inequality is more strict than the 10\% method for $s_0$
providing that $s_0$ is not too large. However, the constraint on the lower bound of $\tau$ is looser than that from the 50\% method,
the largest proportion of ESC can be 60\%-65\% of total phenomenological contributions.

We can not obtain a fixed sum rule window from FIG.~\ref{fig:srwindow} because $s_0$ is not determined at first, thus in
practice, we will initially ``guess'' an $s_0$, then we will obtain a sum rule window $[\tau_{\textrm{min}},\tau_{\textrm{max}}]$
from FIG.~\ref{fig:srwindow}, where $\tau_{\textrm{min}}$ and $\tau_{\textrm{max}}$ are respectively the lower bound and upper bound of the
allowed $\tau$ region with the guessed $s_0$. After finishing the analysis by using this sum rule window, we can check our
initial choice on $s_0$ with its fitted value and adjust it iteratively until it is consistent with the results of the analysis. After finishing
this iterative procedure, the sum rule window is completely determined, all outputs will be obtained self-consistently from the final sum rule window.

\begin{table}[htbp]
\centering
\begin{tabular}{cccccc}
  \hline
  \hline
   & include $\Pi^\alpha_{d=4}$ & $\chi^2/10^{-4}$& $s_0/\textrm{GeV}^2$ & $m_\rho/\textrm{GeV}$ & $\Gamma_\rho/\textrm{GeV}$ \\
  \hline
 \vspace{3pt}
  \multirow{2}{*}{$\kappa=2$} & Yes &1.1& $1.37^{+0.12}_{-0.13}$ & $0.671^{+0.044}_{-0.056}$ & $0.102^{+0.012}_{-0.015}$ \\
   & No &2.1 & $1.57^{+0.10}_{-0.11}$ & $0.739^{+0.032}_{-0.038}$ & $0.119^{+0.009}_{-0.010}$ \\
   \hline
 \vspace{3pt}
 \multirow{2}{*}{$\kappa=3$} & Yes &3.6 & $1.77^{+0.12}_{-0.13}$ & $0.793^{+0.037}_{-0.045}$ & $0.131^{+0.010}_{-0.012}$\\
   & No &5.1& $1.93^{+0.10}_{-0.11}$ & $0.839^{+0.028}_{-0.034}$ & $0.142^{+0.008}_{-0.009}$ \\
   \hline
 \vspace{3pt}
  \multirow{2}{*}{$\kappa=4$} & Yes &7.9& $2.09^{+0.12}_{-0.14}$ & $0.874^{+0.033}_{-0.040}$ & $0.149^{+0.009}_{-0.011}$ \\
  & No &13& $2.23^{+0.11}_{-0.12}$ & $0.911^{+0.027}_{-0.033}$ & $0.158^{+0.008}_{-0.009}$ \\
  \hline
  \hline
\end{tabular}
\caption{\label{tab:fitresult} Fitted results with $\kappa=2,3,4$. All uncertainties of QCD input parameters are set to 10\%.}
\end{table}

The uncertainties of all input parameters in \eqref{eq:input} are set to 10\% which is a typical uncertainty in QCDSR \cite{Leinweber:1995fn}.
After finishing the fitting procedure, we
obtain the results which are listed in TABLE \ref{tab:fitresult}, where the median and the asymmetric standard deviations
from the median for all output parameters are reported. All results are the final statistical results from 2000 fitting
samples, and we report both the results with and without $\alpha_s$ corrections to $\langle \alpha_s G^2\rangle$ and
$\langle m_q\bar qq\rangle$.

From TABLE \ref{tab:fitresult} we find that the uncertainties of phenomenological output parameters are less than 10\%
for $\kappa=3,4$. Only with $\kappa=2$, the uncertainty of $\Gamma_\rho$ will reach to about 15\%, however, even in this
case, the uncertainties of $s_0$ and $m_\rho$ are still less than 10\%. These results imply that the fitted results are
very stable with different input parameters. After including the $\alpha_s$ corrections, we find all values of output parameters
reduce about 4\%-6\% for $\kappa=4$, for small $\kappa$, the reduction is even more apparent. This result is in contrast with the
result in Ref.~\cite{Surguladze:1990sp} where the $\alpha_s$ corrections increased the predicted value of the $\rho$ meson mass.

We also find besides extending the allowed region by the H\"older inequality and reducing the output parameters,
the $\alpha_s$ correction also reduce the value of $\chi^2$.  For a least-squares fitting, a smaller $\chi^2$ means a better
fitting, thus the fitted results are more reliable if these $\alpha_s$ corrections to dimension-4 operators are included.

\begin{figure}[htbp]
\centering
\subfigure[~$\kappa=3$]{\label{fig:scattermass}
\includegraphics[scale=0.9]{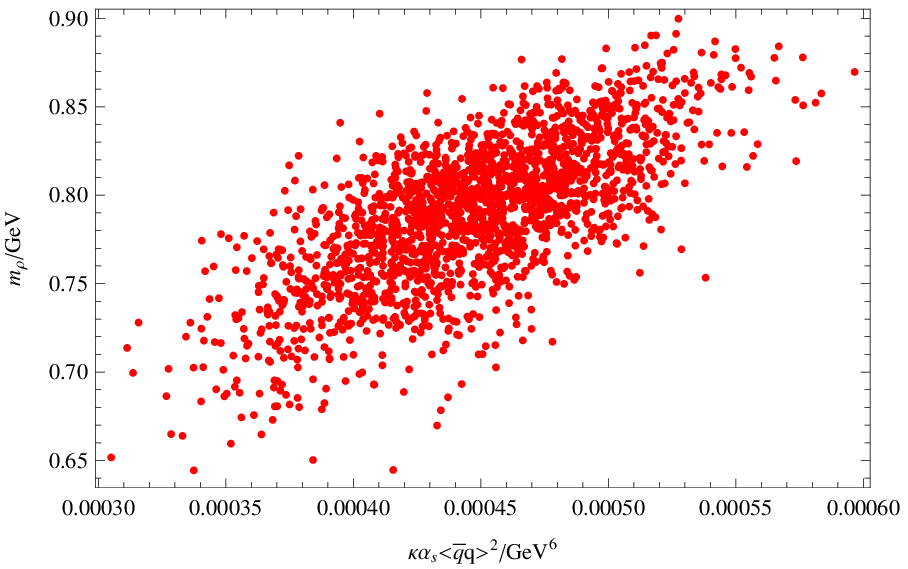}}
\subfigure[~$\kappa=3$]{\label{fig:scattermasswidth}
\includegraphics[scale=0.9]{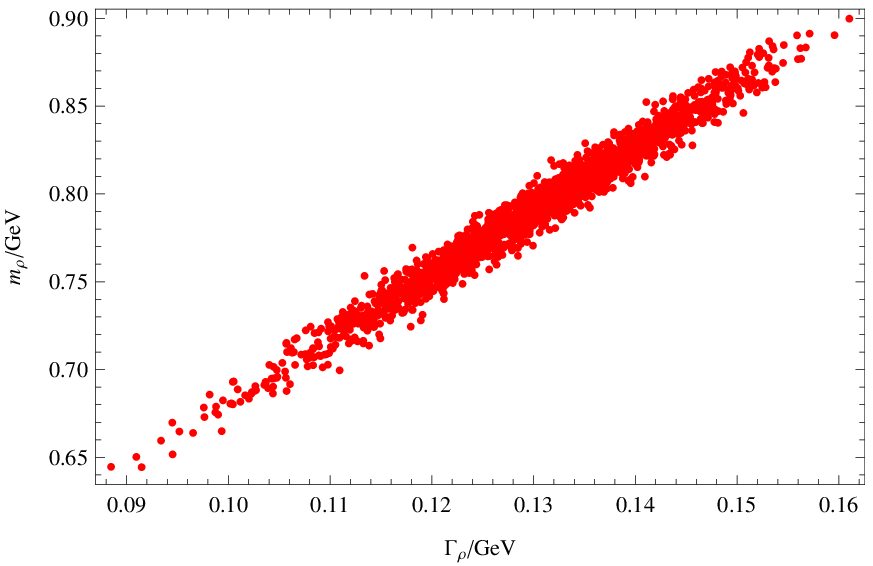}}
\caption{The scatter plot of $m_\rho$ and $\kappa\alpha_s\langle \bar qq\rangle^2$ and the scatter plot of  $m_\rho$ and $\Gamma_\rho$.}
\end{figure}

Based on the 2000 fitting samples, the correlations of all parameters can also be obtained. From the scatter plot
FIG.~\ref{fig:scattermass}, we find that there exists very strong positive correlation between $m_\rho$ and
$\kappa\alpha_s\langle \bar qq\rangle^2$, which means a larger contribution from dimension-6 operator condensate
will lead to a larger $m_\rho$. From FIG.~\ref{fig:scattermasswidth}, we also learn that a larger $m_\rho$ is accompanied
with a larger $\Gamma_\rho$, thus the determination of an exact value of dimension-6 operator condensate is extremely
important. We also find that there exists weak negative correlation between $\langle \alpha_s G^2\rangle$ and all output
parameters, thus any correction to $\langle \alpha_s G^2\rangle$ is important. However, because of the small order of
magnitude, $\langle m_q\bar qq\rangle$ is not important in the sum rules for the $\rho$ meson. Finally, a larger
$\Lambda_{\textrm{QCD}}$ also leads to smaller output parameters, there exists weak negative correlation between them.

From TABLE \ref{tab:fitresult} we find if we choose $\kappa\sim 3$, the fitted result (with 10\% uncertainties on
input parameters) will cover the physical mass of $\rho$ \cite{Olive:2016xmw}. If we want to find an exact match
between the predicted mass and the physical mass, then $\kappa=2.8$ is the best choice, which leads to the result as follows
\begin{equation}
\label{eq:best}
s_0=1.71^{+0.12}_{-0.13}\,\textrm{GeV}^2,~~ m_\rho=0.774^{+0.037}_{-0.046}\,\textrm{GeV},~~\Gamma_\rho=0.126^{+0.010}_{-0.012}\,\textrm{GeV}.
\end{equation}
However, the predicted width is lower than the physical value \cite{Olive:2016xmw} by 16\%, i.e., we
meet a similar problem as Ref.~\cite{DMP}. However, we can conclude from the scatter plot between $m_\rho$ and
$\Gamma_\rho$ that it can not be explained by the uncertainty in the value of the four-quark condensate because if
the width matches the experimental value, then the mass will be too large.

\begin{figure}[htbp]
\centering
\includegraphics[scale=0.9]{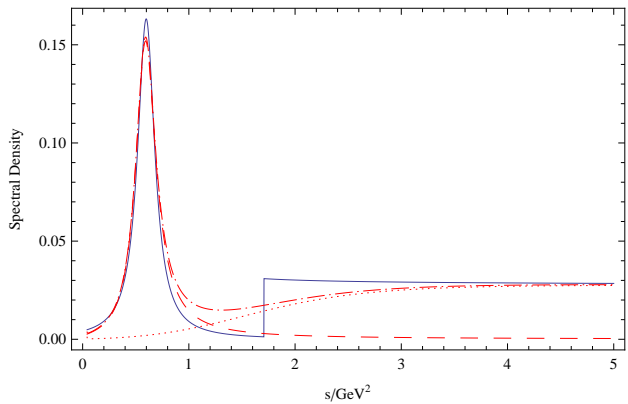}
\caption{\label{fig:spectral} The spectral density for vector $I=1$ channel. The dot-dashed line shows the spectral
density from Eq.~\eqref{eq:shuryak} ($m_\rho$ and $\Gamma_\rho$ are taken from PDG \cite{Olive:2016xmw}) while the
solid line shows the result from our spectral density model \eqref{eq:model} with the median fitted values of
$s_0$, $m_\rho$ and $\Gamma_\rho$ with $\kappa=2.8$. The dashed line and dotted line shows the contribution from the first term
and second term in Eq.~\eqref{eq:shuryak} respectively.}
\end{figure}

Shuryak once parameterized the spectral density from the experimental data for the vector $I=1$ channel by the following function \cite{Shuryak:1993kg}
\begin{equation}
\label{eq:shuryak}
\frac{1}{\pi}{\rm Im}\Pi_\rho (s)=\frac{3}{2\pi^2} \frac{1}{1+4(\sqrt{s}-m_\rho)^2/\Gamma_\rho^2}+\frac{1}{4\pi^2}
\left(1+\frac{\bar\alpha_s(s)}{\pi}\right)\frac{1}{1+e^{(E_0-\sqrt{s})/\delta}},
\end{equation}
where $E_0=1.3\,\textrm{GeV}$, $\delta=0.2\,\textrm{GeV}$ and $\bar \alpha_s(s)=0.7/\ln(\sqrt{s}/0.2\,\textrm{GeV})$.
This function includes a resonance peak (the first term of Eq.~\eqref{eq:shuryak}) and excited states and continuum
contributions (the second term of Eq.~\eqref{eq:shuryak}) which smoothly transition to the perturbative result of the spectral density  as $s$ increases.
Now we can compare our spectral density model with Eq.~\eqref{eq:shuryak}. From FIG.~\ref{fig:spectral}, we can conclude
that the simple spectral density we are using has characterized the main behavior of $\rho$ resonance peak. However, we
also notice that the region around $s_0$ is the main incompatible region in the spectral density, although the value of $s_0$
with $\kappa=2.8$ is very close to $E_0^2$, the ``step'' at $s_0$ in the spectral density is unnatural. A smoothly varying spectral
density should be used in this region, and we conjecture that a better description of the ESC may lead to a better prediction for
$\Gamma_\rho$.

To investigate our conjecture, we replace the $\theta$ function in the traditional ESC model $\frac{1}{\pi}
\textrm{Im}\Pi^{\textrm{(OPE)}}(s)\theta(s-s_0)$ with a smooth function\footnote{We choose the present smooth function
because it includes the needed asymptotic behavior and it is easy to handle in the dispersion integral.}
to construct a new ESC model for the phenomenological spectral density as follows
\begin{equation}
\label{eq:model2}
\frac{1}{\pi}{\rm Im}\Pi^{(\textrm{ESC})}(s)=\frac{1}{4\pi^2}\left(1+\frac{\alpha_s}{\pi}\right)
\cdot\frac{1}{2}\left(\tanh\left(\frac{s-s_0}{\lambda}\right)+1\right),
\end{equation}
where we still associate $s_0$ with the continuum threshold while $\lambda$ ($\geq 0\,\textrm{GeV}^2$) as the parameter
which controls the transition of $\frac{1}{\pi}\textrm{Im}\Pi^{(\textrm{ESC})}(s)$ to the QCD continuum. Obviously,
when $\lambda \to 0\,\textrm{GeV}^2$, the new ESC model will recover the same
ESC as in Eq.~\eqref{eq:model}. Conversely, a larger $\lambda$ will lead to a flatter ESC contribution in
the phenomenological spectral density.

\begin{figure}[htbp]
\centering
\includegraphics[scale=0.9]{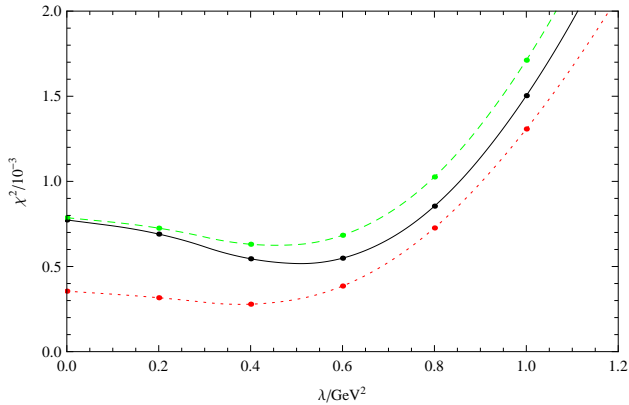}
\caption{\label{fig:plotchi2} The fitted value of $\chi^2$ for selected $\lambda$.
The dots joined with (red) dotted line show the values of $\chi^2$ with $\kappa=3.0$, while
the dots joined by (black) solid line and the
dots joined by (green) dashed line show the values of $\chi^2$ with $\kappa=3.5$ and  $\kappa=4.0$ respectively.
All $\chi^2$ fits with non-zero $\lambda$ use the sum rule window determined by the
$\lambda=0$ H\"older inequality.
}
\end{figure}

To obtain an impression of how the fit is affected by $\lambda$,
we first fix the sum rule window from the $\lambda=0$ H\"older inequality
constraint, then perform the least-$\chi^2$ fit \eqref{eq:chi2} for selected values of
$\lambda$. 
Because all fits (with same $\kappa$) use the same sum rule window, 
we can identify the preferred range of $\lambda$ from values of $\chi^2$.
In Fig.~\ref{fig:plotchi2}, we plot this minimized $\chi^2$ as a function of
$\lambda$ for $\kappa=3.0, 3.5, 4.0$. Although the evidence for a non-zero optimum
value of $\lambda$ is marginal, the rapid increase in $\chi^2$ for large $\lambda$
provides an upper bound of approximately $\lambda\lesssim 1 \,{\rm GeV^2}$.


\begin{figure}[htbp]
\centering
\subfigure[~$\kappa=3.0$]{\label{fig:mwk3}\includegraphics[scale=0.9]{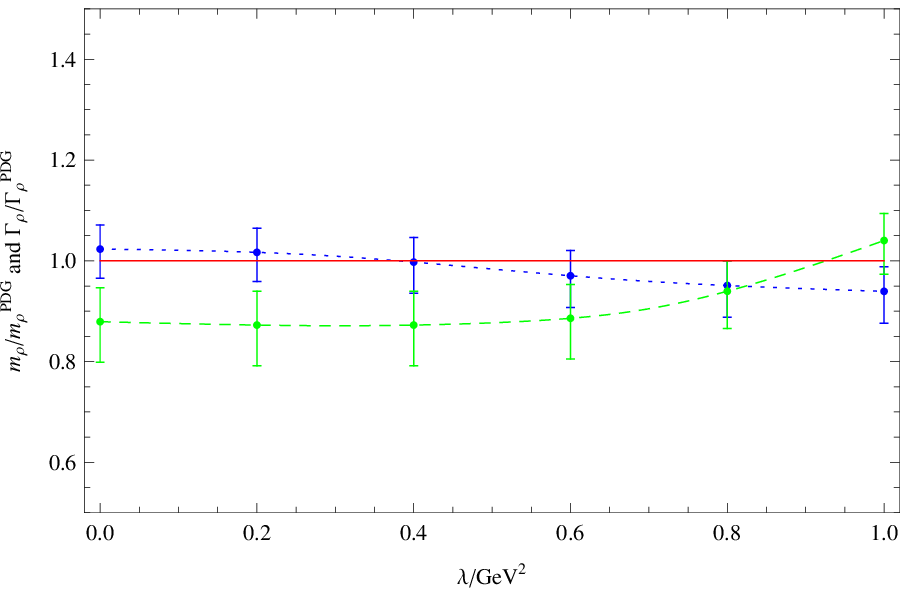}}
\subfigure[~$\kappa=3.5$]{\label{fig:mwk35}\includegraphics[scale=0.9]{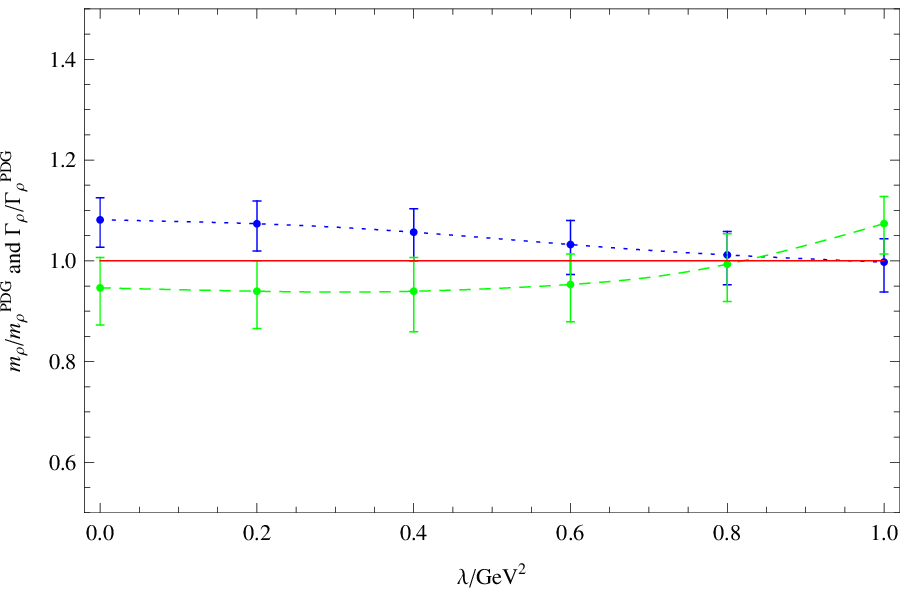}}
\caption{\label{fig:mw} The ratios of $m_\rho/m_\rho^{\textrm{PDG}}$
(joined by (blue) dotted line) and  $\Gamma_\rho/\Gamma_\rho^{\textrm{PDG}}$ (joined by (green) dashed line) versus $\lambda$, where $m_\rho$
and $\Gamma_\rho$ are the median fitted mass and width of $\rho$ meson while
$m_\rho^{\textrm{PDG}}$ and $\Gamma_\rho^{\textrm{PDG}}$ are the mass
and width of $\rho$ meson listed in Review of Particle Physics \cite{Olive:2016xmw}.
The error bars show the uncertainties of the fitted results.
All uncertainties of QCD input parameters are set to 10\%.
}
\end{figure}

\begin{table}[htbp]
\centering
\begin{tabular}{cccccccccc}
  \hline
  \hline
 &\multicolumn{3}{c}{$\kappa=3.0$}&\multicolumn{3}{c}{$\kappa=3.5$}&\multicolumn{3}{c}{$\kappa=4.0$}\\
\hline
  \vspace{3pt}
&$s_0$/GeV$^2$ &$m_\rho$/GeV & $\Gamma_\rho$/GeV&$s_0$/GeV$^2$ &$m_\rho$/GeV & $\Gamma_\rho$/GeV&$s_0$/GeV$^2$ &$m_\rho$/GeV & $\Gamma_\rho$/GeV\\
\hline
  \vspace{3pt}
$\lambda=0.2$\,GeV$^2$ & $1.75^{+0.12}_{-0.14}$ & $0.788^{+0.037}_{-0.045}$ & $0.130^{+0.010}_{-0.012}$ & $1.91^{+0.12}_{-0.14}$ & $0.832^{+0.035}_{-0.042}$
& $0.140^{+0.009}_{-0.011}$ & $2.06^{+0.12}_{-0.14}$ & $0.869^{+0.033}_{-0.041}$ & $0.149^{+0.009}_{-0.011}$\\
\hline
  \vspace{3pt}
$\lambda=0.4$\,GeV$^2$ & $1.65^{+0.13}_{-0.14}$ & $0.773^{+0.038}_{-0.048}$ & $0.130^{+0.010}_{-0.012}$ & $1.82^{+0.13}_{-0.14}$ & $0.819^{+0.036}_{-0.044}$
& $0.140^{+0.010}_{-0.012}$ & $1.97^{+0.13}_{-0.15}$ & $0.857^{+0.034}_{-0.042}$ & $0.148^{+0.009}_{-0.011}$\\
\hline
  \vspace{3pt}
$\lambda=0.6$\,GeV$^2$ & $1.50^{+0.13}_{-0.15}$ & $0.752^{+0.039}_{-0.049}$ & $0.132^{+0.010}_{-0.012}$ & $1.69^{+0.14}_{-0.16}$ & $0.800^{+0.037}_{-0.046}$
& $0.142^{+0.009}_{-0.011}$ & $1.85^{+0.13}_{-0.15}$ & $0.840^{+0.035}_{-0.043}$ & $0.150^{+0.009}_{-0.011}$\\
\hline
  \vspace{3pt}
$\lambda=0.8$\,GeV$^2$ & $1.32^{+0.14}_{-0.16}$ & $0.737^{+0.038}_{-0.049}$ & $0.140^{+0.009}_{-0.011}$ & $1.51^{+0.14}_{-0.16}$ & $0.784^{+0.036}_{-0.046}$
& $0.148^{+0.009}_{-0.011}$ & $1.68^{+0.14}_{-0.16}$ & $0.823^{+0.035}_{-0.044}$ & $0.154^{+0.009}_{-0.010}$\\
\hline
  \vspace{3pt}
$\lambda=1.0$\,GeV$^2$ & $1.11^{+0.15}_{-0.18}$ & $0.728^{+0.038}_{-0.049}$ & $0.155^{+0.008}_{-0.010}$ & $1.31^{+0.15}_{-0.17}$ & $0.773^{+0.036}_{-0.046}$
& $0.160^{+0.008}_{-0.009}$ & $1.47^{+0.14}_{-0.17}$ & $0.809^{+0.034}_{-0.043}$ & $0.165^{+0.008}_{-0.009}$\\
 \hline
  \hline
\end{tabular}
\caption{\label{tab:fitresult2} Fitted results with $\kappa=3.0,3.5,4.0$ and $\lambda=0.2-1.0\,\textrm{GeV}^2$.
All uncertainties of QCD input parameters are set to 10\%.
}
\end{table}

By setting $\lambda=0.2, 0.4, 0.6, 0.8, 1.0\,\textrm{GeV}^2$, we can obtain the allowed ($\tau$, $s_0$) 
regions from the H\"older inequality respectively, then the iterative fitting procedure
introduced in the previous section can be invoked.
We plot ratios of the fitted mass and width to the experimental
values as $\lambda$ increases in Fig.~\ref{fig:mw} (related detailed fitted results are listed in
Table \ref{tab:fitresult2}). From Fig.~\ref{fig:mw} and Table \ref{tab:fitresult2},
we find that a larger value of $\lambda$ will generally lead to a smaller $m_\rho$,
and a larger $\Gamma_\rho$ (the fitted $\Gamma_\rho$ is not sensitive with $\lambda$
with $\lambda<0.4\,\textrm{GeV}^2$), furthermore, increasing $\kappa$, increases both
$m_\rho$ and $\Gamma_\rho$. These tendencies demonstrate that with an appropriately
smoothed ESC and $0.4\,{\rm GeV^2}<\lambda<1\,{\rm GeV^2}$, we may obtain results for
 $\Gamma_\rho$ and $m_\rho$ statistically consistent with their experimental values.

\begin{figure}[htbp]
\centering
\includegraphics[scale=0.9]{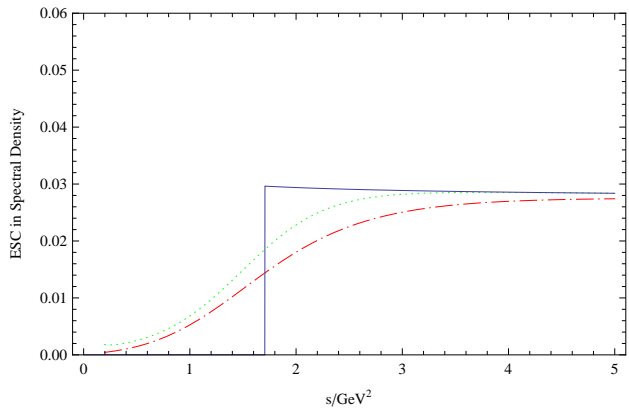}
\caption{\label{fig:esc} The ESC in the spectral density for vector $I=1$ channel.
The (red) dot-dashed line shows the ESC from the second term in Eq.~\eqref{eq:shuryak}, the
(green) dotted line shows the result from our ESC model \eqref{eq:model2} with $\kappa=3.5$
and $\lambda=0.8\,\textrm{GeV}^2$, and the (blue) solid line shows
the ESC from the phenomenological spectral density \eqref{eq:model} with $\kappa=2.8$.}
\end{figure}

To obtain an intuitive impression on the behavior of the ESC contributions,
we plot the two spectral density models used in this paper in Fig.~\ref{fig:esc}, where
we also plot the ESC from Shuryak's spectral density \eqref{eq:shuryak}. From this figure, we find
that our new ESC model describes the main behavior of Shuryak's ESC \cite{Shuryak:1993kg} better than the
traditional ESC model.


Finally, as a comparison, we also try the fitting procedure with the traditional phenomenological narrow width spectral density
\begin{equation}
\frac{1}{\pi}{\rm Im}\Pi^{\textrm{(phen)}}= f^2_\rho \delta(s-m_\rho^2)+\frac{1}{4\pi^2}\left(1+\frac{\alpha_s}{\pi}\right)\theta(s-s_0),
\end{equation}
where $f_\rho$ is the coupling constant of $\rho$ meson.
However, the outputs of $s_0$ with $\kappa=\text{2--4}$ are too small, thus the sum rule window departs from the allowed region from the H\"older
inequality. Even without violation of factorization, we still can not find a result in this procedure which is completely consistent
with the H\"older inequality.

\section{Summary and Discussion}

In this paper, we have introduced a systematic sum rule approach to give predictions for the hadronic mass and width.
This approach is a synthesis of the Monte-Carlo based QCD sum rule and the H\"older-inequality-determined
sum rule window, accompanied with a phenomenological spectral density of Breit-Wigner form that is normalized by the low-energy theorem. We apply this
approach to the $\rho$ meson to re-analyze its phenomenological properties and the corresponding theoretical uncertainties.
We also calculate the $\alpha_s$ corrections to dimension-4 condensates in the OPE of light vector $I=1$ current
correlation function by using the external field method in Feynman gauge. Our calculation confirm the previous results
obtained using the projector method \cite{Surguladze:1990sp}, which have not previously been
considered in the existing sum rule analyses. Considering these higher
order effects in the perturbation series, we conducted a comprehensive reanalysis with the new methodological
approach. The findings of our numerical analysis are:
\begin{itemize}
\item The Breit-Wigner type spectral density model, which provides a better description of physical spectral density
than the traditional single narrow resonance spectral model, plays an important role in our procedure. First,
the sum rule window can be completely determined by the H\"older inequality in this spectral density model, thus we avoid the
often criticized 50\%-10\% assumptions used to constrain the sum rule window in previous Monte-Carlo based
QCD sum rules analysis \cite{Leinweber:1995fn,Erkol:2008gp,Zhang:2013rya}.
Meanwhile, the traditional single narrow resonance phenomenological spectral density is too over-simplified to find a fitted result consistent
with the H\"older inequality. Second, because of the low-energy theorem normalization for the electromagnetic form factor in our phenomenological spectral density,
the degrees of freedom during the fitting procedure reduces to three phenomenological output parameters, i.e., $s_0$, $m_\rho$ and $\Gamma_\rho$
and improves the ability of QCD sum rules to predict the width of $\rho$ meson.
The three-parameter model is a clear improvement on the unphysical $\Gamma_\rho=0$ result \cite{Zhang:2013rya} that is found in the
four-parameter spectral density model $\frac{1}{\pi}{\rm Im}\Pi(s)= f_\rho^2 \frac{1}{\pi}
\frac{m_\rho\Gamma_\rho}{(s-m_\rho^2)^2+m_\rho^2\Gamma_\rho^2}+\frac{1}{\pi}{\rm Im}\Pi^{\textrm{(OPE)}}(s)\theta(s-s_0)$.
\item Although most works in the literature do not consider the effect of $\alpha_s$ corrections to dimension-4 operators,
we find they play an important role in the present QCD sum rules analysis for $\rho$ meson by extending the $s_0$-$\tau$ region allowed by the H\"older
inequality and improving the goodness for fit between the theoretical side and the phenomenological side of the master equation for
QCD sum rules. These radiative corrections also reduce the values of $s_0$, $m_\rho$ and $\Gamma_\rho$ by 4\%-16\%, depending on the value of $\kappa$.
Thus it is important to include these corrections in our QCDSR approach, although the order of magnitude of these corrections is small.
\item With 10\% uncertainties for input parameters, the optimal phenomenological results in our numerical analysis are:
$m_\rho=0.774^{+0.037}_{-0.046}\,\textrm{GeV}$, $\Gamma_\rho=0.126^{+0.010}_{-0.012}\,\textrm{GeV}$ and
$s_0=1.71^{+0.12}_{-0.13}\,\textrm{GeV}^2$ with $\kappa=2.8$ from the traditional ESC model.
All the outputs are consistent with the experimental values for the $\rho$ meson. The factorization violation
factor $\kappa\sim3$ is also the expected size estimated from many different channels.
Extending our phenomenological model to include a smooth transition to the QCD continuum (similar to Ref.~\cite{Shuryak:1993kg})
leads to generally better agreement with the experimental values.
\end{itemize}

The successful application of our approach in the $\rho$ channel greatly motivates us to extend it to other channels.
Since similar phenomenological spectral densities can be constructed by inserting a complete set of intermediate states
for other hadrons which have dominant decay modes, such extensions are expected to be applicable. For example, by inserting
two-pion intermediate states in the correlation function for $I=0$ scalar current, we can connect the phenomenological
spectral density with scalar pion form factor $F_s(s)$. Then the result of $F_s(0)=m_\pi^2$ in chiral perturbation theory \cite{Gasser:1990bv}
can be used to reduce the number of the phenomenological parameters to three.
Furthermore, the H\"older inequality constraint to determine the sum rule window of validity is similarly adaptable to other channels.
Subsequent works can examine the validity of our new method and help us further improve it.

\begin{acknowledgments}
This work is supported by NSFC under grant 11175153, 11205093 and 11347020, and supported by Open Foundation of the Most Important
Subjects of Zhejiang Province, and K. C. Wong Magna Fund in Ningbo University.
TGS is supported by the Natural Sciences and Engineering Research Council of Canada (NSERC). Z. F. Zhang and Z. R. Huang are
grateful to the University of Saskatchewan for its warm hospitality.
\end{acknowledgments}

\appendix

\section*{Appendix}

We calculate the $\alpha_s$ corrections to $\langle\alpha_{s}G^2\rangle$ and $\langle m_{q}\bar{q}q\rangle$ by using external
field method in Feynman gauge in space-time dimension $d=4-2\epsilon$.

\begin{figure}[htbp]
\centering
\includegraphics[scale=0.9]{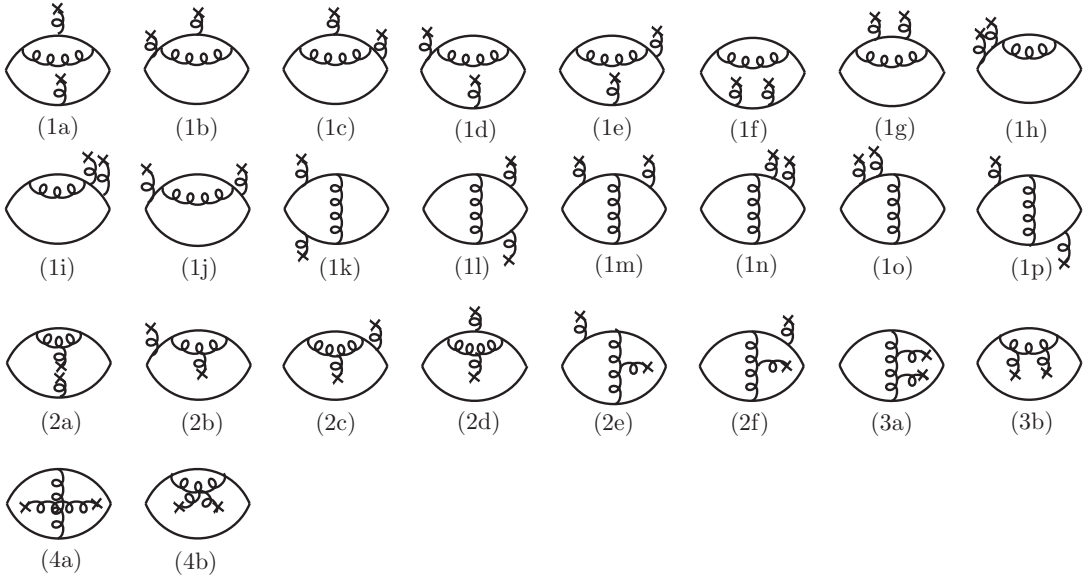}
\caption{\label{fig:gg} Feynman diagrams for $\alpha_s$ corrections to $\langle\alpha_s G^2\rangle$.}
\end{figure}

The two-loop diagrams  which contribute to $\alpha_s$ corrections to $\langle \alpha_s G^2\rangle$ are shown in FIG.~\ref{fig:gg}.
The final result of the $\alpha_s$ correction to $\langle \alpha_s G^2\rangle$ reads
\begin{equation}
\label{eq:gg}
\Pi_{\langle\alpha_s G^2\rangle}(q^{2})=\frac{1}{(q^{2})^{2}}\frac{7}{72\pi}\frac{\alpha_{s}}{\pi}\langle\alpha_{s}G^2\rangle,
\end{equation}
where we have taken the mass of light quark to be zero. This result is consistent with the result obtained in Ref.~\cite{Surguladze:1990sp}
from a different calculation method. Explicit results for the diagrams in FIG.~\ref{fig:gg} are listed in TABLE \ref{tab:aggfull},
where we divide the results into several parts according to the diagrams with different kinds of interaction vertices.

\begin{figure}[htbp]
\centering
\includegraphics[scale=0.9]{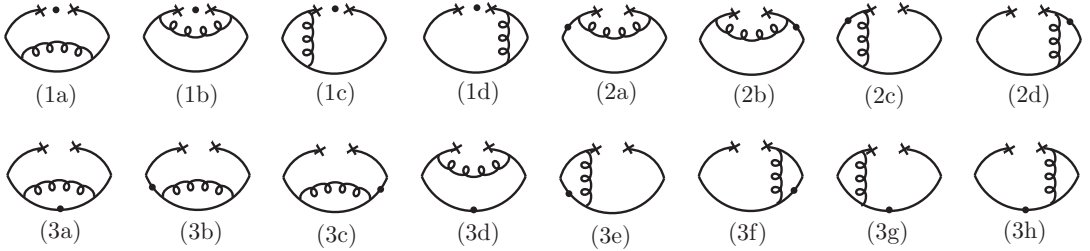}
\caption{\label{fig:qq} Feynman diagrams for $\alpha_s$ corrections to $\langle m_q \bar qq\rangle$.  Dots stand for light quark
mass $m_{q}$ ($m_u=m_d\equiv m_q$).}
\end{figure}

Feynman diagrams which contribute to $\alpha_s$ corrections to $\langle m_q \bar qq\rangle$ are shown in FIG.~\ref{fig:qq}.
After some calculations, we obtain the final result of $\alpha_s$ correction to $\langle m_q\bar qq\rangle$ as
\begin{equation}
\label{eq:mqq}
\Pi_{\langle m_q\bar{q}q\rangle}(q^{2})=\frac{1}{(q^{2})^{2}}\frac{2}{3}\frac{\alpha_{s}}{\pi}\langle m_{q}\bar{q}q\rangle,
\end{equation}
where we have set the masses of light quarks to be equal for convenience, i.e., $m_u=m_d=m_q$.
The explicit results for the diagrams in FIG.~\ref{fig:qq} are listed in TABLE \ref{tab:mqqfull}.

\begin{table}[htbp]
\centering
\begin{tabular}{ll}
  \hline
  \hline
  Diagrams & Coefficients for $\frac{\alpha_s}{(q^2)^2\pi}\langle\alpha_s G^2\rangle$ \\
  \hline
  FIG.~\ref{fig:gg}: (1a)-(1p)& $-\frac{1}{16\pi \epsilon }+\frac{1}{8\pi }\ln \left(\frac{-q^{2}}{\mu^{2}}\right) -\frac{61}{144\pi }+\frac{\gamma_{E}}{8\pi}-\frac{\ln(4\pi)}{8\pi}+\frac{\zeta(3)}{8\pi}$ \\
  FIG.~\ref{fig:gg}: (2a)-(2f) & $\frac{5}{16\pi\epsilon }-\frac{5}{8\pi }\ln \left(\frac{-q^{2}}{\mu^{2}}\right)+\frac{5}{24\pi}-\frac{5\gamma_{E}}{8\pi}+\frac{5\ln(4\pi)}{8\pi}+\frac{3\zeta(3)}{4\pi}$ \\
  FIG.~\ref{fig:gg}: (3a)-(3b) & $-\frac{11}{32\pi\epsilon }+\frac{11}{16\pi }\ln \left(\frac{-q^{2}}{\mu^{2}}\right)+\frac{11}{64\pi}+\frac{11\gamma_{E}}{16\pi}-\frac{11\ln(4\pi)}{16\pi}-\frac{7\zeta(3)}{8\pi}$ \\
  FIG.~\ref{fig:gg}: (4a)-(4b) & $\frac{3}{32\pi\epsilon }-\frac{3}{16\pi }\ln \left(\frac{-q^{2}}{\mu^{2}}\right)+\frac{9}{64\pi}-\frac{3\gamma_{E}}{16\pi}+\frac{3\ln(4\pi)}{16\pi}$ \\
  \hline
  \hline
\end{tabular}
\caption{\label{tab:aggfull}Detail results of $\alpha_s$ corrections to $\langle\alpha_s G^2\rangle$.}
\end{table}

\begin{table}[htbp]
\centering
\begin{tabular}{ll}
  \hline
  \hline
  Diagrams & Coefficients for $\langle m_q \bar qq\rangle$ \\
  \hline
FIG.~\ref{fig:qq}: (1a)-(1d) & $\frac{1}{(q^{2})^{2}}\frac{2}{3}\frac{\alpha_{s}}{\pi}$\\
FIG.~\ref{fig:qq}: (2a)-(2d) & $-\frac{1}{(q^{2})^{2}}\frac{8}{3}\frac{\alpha_{s}}{\pi}$\\
FIG.~\ref{fig:qq}: (3a)-(3h) & $\frac{1}{(q^{2})^{2}}\frac{8}{3}\frac{\alpha_{s}}{\pi}$\\
  \hline
  \hline
\end{tabular}
\caption{\label{tab:mqqfull}Detail results of $\alpha_s$ corrections to $\langle m_q \bar qq\rangle$.}
\end{table}


\end{document}